\documentclass[11pt]{cernrep} 
\usepackage{epsfig} 

\begin{document}

\renewcommand{\baselinestretch}{0.9}
\newcommand{\ccaption}[2]{
    \begin{center}
    \parbox{0.85\textwidth}{
      \caption[#1]{\small{{#2}}}
      }
    \end{center}
    }

\newcommand     \TR  {TopReX } 
\newcommand     \PYT {PYTHIA } 
\newcommand     \TTF {\ttfamily} 

%   NTUPLE

%draw box around routine or common block name

\setlength{\fboxrule}{1pt}
\setlength{\fboxsep}{3mm}
\newcommand{\drawbox}[1]{\vspace{\baselineskip}\noindent%
\fbox{\texttt{#1}}\vspace{0.5\baselineskip}}
\newcommand{\itemc}[1]{\item[\textbf{#1}\hfill]}
\newcommand{\iteme}[1]{\item[\texttt{#1}\hfill]}
\newcommand{\itemn}[1]{\item[{#1}\hfill]}

\newenvironment{entry}%
{\begin{list}{}{\setlength{\topsep}{0mm} \setlength{\itemsep}{0mm}
\setlength{\parskip}{0mm} \setlength{\parsep}{0mm}
\setlength{\leftmargin}{20mm} \setlength{\rightmargin}{0mm}
\setlength{\labelwidth}{18mm} \setlength{\labelsep}{2mm}}}%
{\end{list}}%
\newenvironment{subentry}%
{\begin{list}{}{\setlength{\topsep}{0mm} \setlength{\itemsep}{0mm}
\setlength{\parskip}{0mm} \setlength{\parsep}{0mm}
\setlength{\leftmargin}{10mm} \setlength{\rightmargin}{0mm}
\setlength{\labelwidth}{18mm} \setlength{\labelsep}{2mm}}}%
{\end{list}}

\begin{titlepage}
\nopagebreak
%\thispagestyle{empty}
%\vspace*{2cm}
 \begin{center}
  \begin{Large} 
{\large \bf \sc T~O~P~R~E~X~ generator (version~3.25) } \\
{\bf \sc short manual }
%\end{center}
\end{Large}

\normalsize
\vfill                                                         
\vskip .5cm
%\vspace*{0.5cm}
%\begin{center}
{\bf S. R.~Slabospitsky}\\
{\it Institute for High Energy Physics, \\ Protvino, Moscow Region,
142284 RUSSIA}\\
\verb+ slabospitsky@mx.ihep.su+\\
\bigskip
{\bf L.~Sonnenschein }\\ 
Formerly: \hspace*{0.5ex} {\it Rheinisch Westf\"alische Technische Hochschule (RWTH) 
Aachen, \\
III. Physikalisches Institut, Lehrstuhl B, 52056 Aachen, GERMANY } \\
Now: \hspace*{0.5ex} {\it Boston University, \\
Physics Department, Boston, MA 02215, USA} \\
\verb+  Lars.Sonnenschein@cern.ch+\\
\verb+  sonne@physics.bu.edu+\\
\end{center}

\nopagebreak
\vfill
\vskip 2cm
\begin{abstract} 
This is the manual of the specialized event generator {\bf TopReX~3.25}. 
The generator provides the simulation of several important processes in $pp$ 
and $p {\bar p}$ collisions, not 
implemented in {\bf\PYT} (yet). Some of these processes include $t$-quarks 
whose spin polarizations 
are taken into account in the subsequent decay of the $t$-quarks.
Several non-SM top quark decay channels
are included, too. All calculated subprocesses can be accessed from \PYT as 
external processes.
In addition, \TR can be used as stand alone event generator, providing 
partonic final states before showering.
In this mode the control of the event generation is taken by \TR itself. 
A few simple examples of {\TTF main} routines, which show how to use \TR in 
the different modes are
discussed.
\end{abstract}                                                          

\vskip 1cm
\today \hfill 
\vfill 
\end{titlepage}

\newpage
\setcounter{page}{1}

%%%%%%%%%%%%%%%%%%%%%%%%%%%%%%%%%%%%%%%%%%%%%%%%%%%%%%%%%%%%%%%%%%%%%%%%%%%%%%
\section*{\bf Introduction }
   
 The event generator {\bf\TR} in its present version ({\sf 3.25}) provides the
 generation of important
heavy particle production processes in $pp$ and $p \bar p$ collisions 
($t$-quarks, charged Higgs, etc.), which are 
not implemented in the high energy physics event generator 
{\bf\PYT}~\cite{pythia} (yet).  
The subprocesses implemented in the present version of \TR are listed in 
Table~\ref{txnump}. 

%\end{document}

\begin{table}[h] 
\begin{center} 
\begin{tabular}{rcll} 
 $ q(\bar q) b$ &$\to$& $q'(\bar q') t     $ & {\TTF IPROC = \phantom{2}1 } \\ 
 $ q(\bar q) g$ &$\to$& $q'(\bar q') t \bar b$&{\TTF IPROC = \phantom{2}2 } \\ 
 $ g b        $ &$\to$& $t W$               & {\TTF IPROC = \phantom{2}3 } \\ 
 $ q  \bar q' $ &$\to$& $W^* \, \to t \bar b $&{\TTF IPROC = \phantom{2}4 } \\
 $ q  \bar q' $ &$\to$ & $W Q \bar Q, \,\, Q=b,c$ &{\TTF IPROC = \phantom{2}5 } \\
 $ q  \bar q' $ &$\to$ & $W^* \to \tau \nu $ &{\TTF IPROC = \phantom{2}6  } \\ 
 $ q  \bar q' $ &$\to$ & $W^* \to \tau \nu + jet$ &{\TTF IPROC = \phantom{2}7 } \\
 $ q  \bar q' $ &$\to$ & $H^{\pm*} \to t \bar b$  &{\TTF IPROC = \phantom{2}8 } \\
 $ q  \bar q' $ &$\to$ & $H^{\pm*} \to \tau \nu$  &{\TTF IPROC = \phantom{2}9  } \\ 
 $ q  \bar q' $ &$\to$ & $H^{\pm*} \to \tau \nu + jet$ &{\TTF IPROC = 10 } \\
% \\[1ex] 
 $ gg\, + \,q \bar q$ &$\to$ & $t \bar t$         & {\TTF IPROC = 20 } \\
 $ gg               $ &$\to$ & $t \bar t$         & {\TTF IPROC = 21 } \\
 $ q \bar q         $ &$\to$ & $t \bar t$         & {\TTF IPROC = 22 } \\
 $ gg \, + \,q \bar q$ &$\to$ & $(H \to) ~~~ t \bar t$ & {\TTF IPROC = 50 } \\
 $ gg $ &$\to$ & $(H \to) ~~~ t \bar t$         & {\TTF IPROC = 51 } \\
 $ q \bar q $ &$\to$ & $(H \to) ~~~ t \bar t$         & {\TTF IPROC = 52 } 
\end{tabular} 
\end{center}
\vspace*{-3ex}
\caption{ 
 \label{txnump}
 \TR processes and corresponding process numbers. 
 The user can choose a specific process by passing, among others,
 the wanted process number {\tt IPROC} to the initialization subroutine 
{\tt TOPREX(FRAME, BEAM, TARGET, IPROC, ECM)}.
}
\end{table} 

The polarization of final $t$-quarks is calculated and taken into account in
 the subsequent decay of the $t$-quarks.
For processes with $\tau$-leptons in the partonic final state (originating 
from top decays or
directly produced by $W$ or $H^{\pm}$ bosons) the polarization of the 
$\tau$-leptons is calculated, too. The decay of the polarized $\tau$-leptons 
is treated by the {\bf TAUOLA} package~\cite{tauola}. 
Thus \TR provides the generation of hard subprocesses as listed
in Table~\ref{txnump} with subsequent decays of heavy particles, given
in Table~\ref{txdec}.

\begin{table}[h] 
\begin{center} 
\begin{tabular}{rcl} 
 $ t $ &$\to$ & $bW^+, \,\, \to b H^+, \,\, \to q\gamma,\,\,qg, \,\, qZ$ \\
 $W,Z    $ &$\to$ & $f \bar f', \,\,\, f = q, l, \nu$             \\
 $H^{\pm}$ &$\to$ & $f \bar f', \,\,\, f = q, l, \nu$             \\
 $ \tau^-$ & $\to$ & $\nu_{\tau} l^- \nu, \,\, \nu_{\tau} \pi^-, ...$ 
              {\sf TAUOLA} 
\end{tabular} 
\end{center}
\vspace*{-3ex}
\caption{ 
 \label{txdec}
 Decay channels of heavy particles included in \TR. }
\end{table} 

\noindent 
After the generation of a hard subprocess, \TR returns for any generated event
the following information: \\
$\bullet$ the differential cross section value, \\
$\bullet$ flavours and momenta of the initial and final state partons, \\
$\bullet$ colour flow information, \\
$\bullet$ parton shower arrangements (shower pairs/partners and scales), \\
which is stored in the \PYT common block {\tt /PYUPPR/}.
Subsequently, \PYT can be used for the fragmentation of quarks and gluons into
 jets, followed by the hadronization and the decay of resonances. 
Initial and final state radiation as well as multiple interaction models can 
be used as usual. In analogy to \PYT internal 
processes all \TR subprocesses can be accessed by
\PYT through the call of an external process (one at a time). 
After the simulation of a hard subprocess in \TR through a \PYT call of the
 {\TTF PYUPEV} subroutine, the standard \PYT common block
{\TTF /PYJETS/} is filled with all necessary information about the flavour 
and momenta of the initial and final state partons, the colour flow and
parton shower arrangements and scales.  

There are two alternative ways in which \TR can be used. Fully integrated 
into \PYT by linkage of the \TR package as matrix element library,
respecting the programming structure of \PYT in such a way that the 
impact for the user has been minimized
or as stand-alone program avoiding the overhead of \PYT if it is not needed.
While the former is more practical for the usage of already existing analysis
routines and interfaced detector simulations, the latter provides a complete
analysis environment with the possibility of booking and filling histograms. 

Before the usage of \TR will be discussed in more detail a closer look at the
data structure and program flow gives useful 
insights in the functionality of the program.

%%%%%%%%%%%%%%%%%%%%%%%%%%%%%%%%%%%%%%%%%%%%%%%%%%%%%%%%%%%%%%%%%%%%%%%%%%%%%
\section{\bf General structure of \TR package } %and Input/Output data } 

\begin{figure}[h]
\begin{center}
%\scalebox{0.92}{ 
\unitlength 10mm
\hspace*{-4mm}
\begin{picture}(15,14.5)

\put(5.5,14.0){Program flow chart}

\vspace*{3mm}

\put(0.4,11.65){Program}
\put(0.0,10.5){\thicklines\framebox(2.5,0.7){MAIN}}
\put(0.0,6.5){\framebox(2.5,4.0){ 
 { \begin{tabular}{c} Initialization \\ {} \\
     \dashbox{0.1}(2.0,1.0) {\begin{tabular} {c} PYTHIA \\ parameters
                 \end{tabular} } \\ {} \\ {} \\ {} 
 \end{tabular}  }  }}

\put(0.0,1.5){\framebox(2.5,5.0){ 
 { \begin{tabular}{c} Event loop \\ {} \\ {} \\ {} \\ {} \\
     \dashbox{0.1}(2.0,1.0) {\begin{tabular} {c} analysis \\ routines 
                 \end{tabular} } \\ {} \\ {} \\ {} 
 \end{tabular}  } }}

\put(0.0,1.0){\framebox(2.5,0.5){End}}

\put(2.5,7.5){\vector(1,0){2.0}}
\put(4.5,7.15){\thicklines\framebox(2.5,0.7){PYINIT}}
%\put(5.7,9.1){\line(0,-1){1.0}}
%\put(5.7,8.1){\vector(1,0){3.8}}

%\multiput(2.5,8.75)(0.5,0.0) {15} {$-$~} 
\put(2.5,9.715){\vector(1,0){7.0}}
\put(9.5,9.4){\thicklines\framebox(2.5,0.7){TOPREX}}
\put(10.7,9.4){\line(0,-1){1.8}} % {4.1}}

\put(10.7,8.6){\vector(1,0){1.5}}
\put(12.2,8.25){\thicklines\framebox(2.5,0.7){TX\_TIT}}
\put(10.7,7.6){\vector(1,0){1.5}}
\put(12.2,7.25){\thicklines\framebox(2.5,0.7){TXWEAK}}

\put(2.5,6.0){\vector(1,0){1.0}}
\put(3.5,5.65){\thicklines\framebox(2.5,0.7){PYEVNT}}
\put(4.7,5.6){\line(0,-1){1.0}}
\put(4.7,4.6){\vector(1,0){1.0}}
\put(5.7,4.25){\thicklines\framebox(2.5,0.7){PYRAND}}
\put(5.5,3.2){\vector(0,-1){1.5}}
\put(4.2,1.2){ {\tt common /PYJETS/} }

%\put(6.9,4.25){\line(0,-1){1.0}}
\put(8.25,4.55){\vector(1,0){2.25}}
\put(10.5,4.25){\thicklines\framebox(2.5,0.7){PYUPEV}}
\put(10.5,3.55){\framebox(2.5,0.7){\scalebox{0.75}{$|{\cal{M}} 
                         |^2\otimes LIPS$}}}
\put(11.5,3.2){\vector(0,-1){0.4}}
\put(10.2,2.5){ {\tt common /PYUPPR/} } 
\put(10.1,2.6){\vector(-1,0){2.5}}

\put(9.1, 0.8){\dashbox{0.2}(6.0,12.0){ } }
\put(10.9,11.6){\Large \sf TopReX}

\put(3.1, 0.8){\dashbox{0.2}(5.4,12.0){ } }
\put(4.75,11.6){\large \sf PYTHIA}

\end{picture}
%} % end scalebox
\end{center}

\caption{ \label{programflowchart}
  Program flow chart. The main program, stored in the physical file main.f, 
initializes \TR and \PYT.
  The initialization of \TR consists predominantly of the passage of 
electroweak parameters from \PYT to \TR followed by the estimation 
  of the maximal value of the differential cross section for the chosen 
hard scattering process. After the initialization
  the generation of scattering events takes place in the event loop of the 
main program through a call to the \PYT subroutine {\TTF PYEVNT}.
  The subroutine {\TTF PYUPEV} (whose \PYT dummy version is overwritten 
by \TR) is called subsequently. 
}

\end{figure}
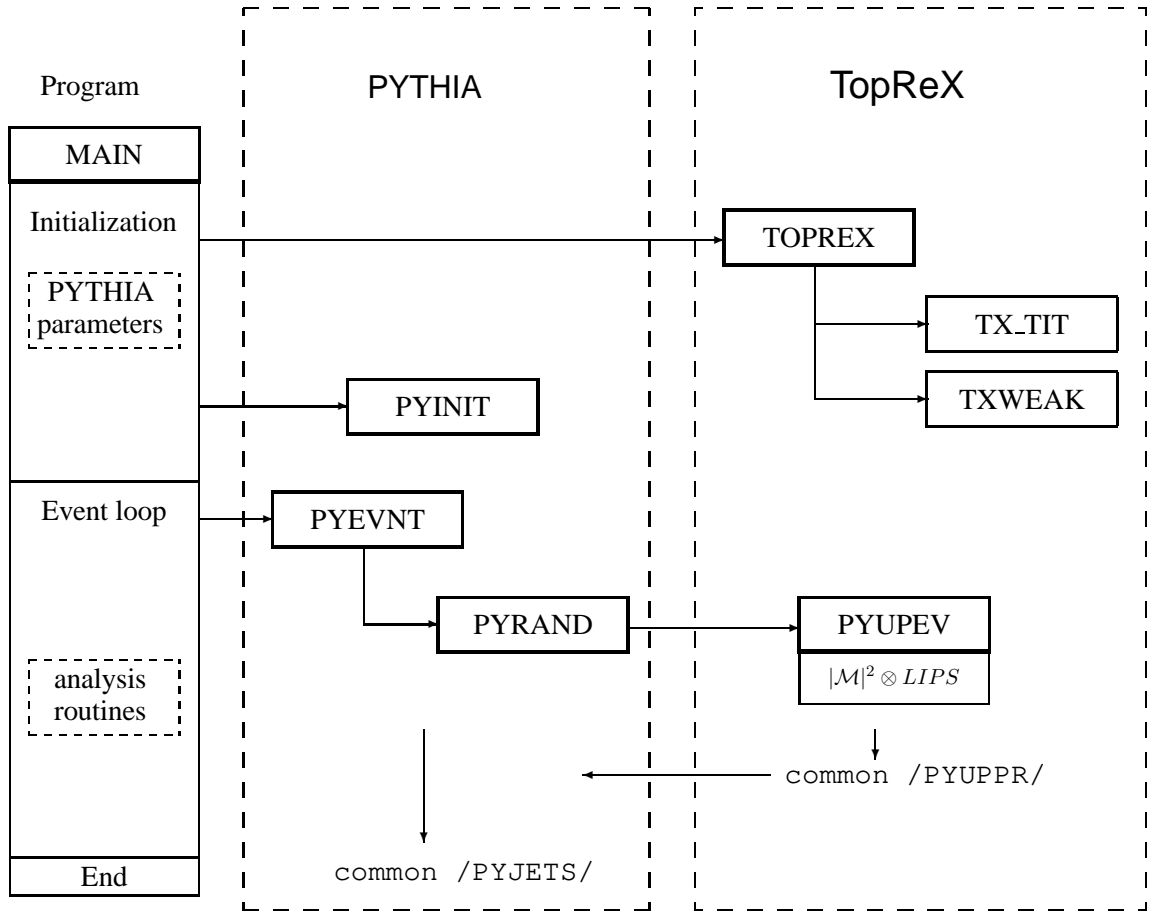

The structure of \TR and the corresponding program flow is
shown in fig.~\ref{programflowchart}.

If needed, the user can modify \PYT parameters first.
This can be done in the {\tt main.f} routine or in a separate user's subroutine
(e.g. subroutine {\tt TXPYIN}, see appendix~2). 
Two examples of {\tt main.f} routines and a
corresponding subroutine to set \PYT parameters are presented in
the appendices~1 and~2. %%%%% \ref{appA} and ~ \ref{appB}. 

Then the subroutine {\tt TOPREX(FRAME, BEAM, TARGET, IPROC, ECM)} has to be 
called, in which 
all variables are identical to those used in the call of the {\tt PYINIT} 
subroutine
except the integer variable {\tt IPROC}, which specifies the number of the 
\TR process.
The character variable {\tt FRAME}  is foreseen to pass the information of 
the rest frame 
in which the generated event has to be evaluated.
The colliding particles 
have to be specified in the character variables {\tt BEAM} and {\tt TARGET} 
and the center of mass energy of the collider has to be given in the variable 
{\tt ECM}.
These are all parameters, needed to run \TR. Only in the case of very specific
 processes with
Flavour Changing Neutral Current (FCNC) interactions the appropriate
parameters should be specified in the subroutine {\tt FCNC\_INI}.

After the initialization of \TR the subroutine {\tt PYINIT} has to be called 
for the initialization of \PYT. 
Subsequently events can be generated. This is typically done in an event loop 
as shown in the example
program {\tt main.f}. At this place the user may call its own analysis 
routines (see appendix~2). 

\PYT can access the processes of \TR as external processes through a call to 
the subroutine {\tt PYUPEV}.
This routine, in turn,  calls the appropriate \TR routines, needed to 
evaluate matrix elements squared, Lorentz invariant phase spaces (LIPS), parton
luminosities (quark and gluon distributions of the beam particles) as well as 
to provide colour flow 
information and to set up parton shower arrangements and scales
of the hard subprocess. Finally, all information of the generated event
is stored in \PYT common block {\TTF /PYUPPR/}. Returned from the 
{\tt PYUPEV} subroutine, \PYT applies its standard
procedure of jet fragmentation, hadronization and resonance decays. The 
simulated events are written to
the standard common block {\TTF /PYJETS/}.

As default, control printout is sent to the terminal. In addition a second
possibility to store the printout in
a separate file ({\tt toprex.info}) is provided. In this case, the user 
should call the subroutine  {\tt TXWRIT(-6)} 
in the {\tt main.f} routine before any initialization. The parameter
value '-6' means, that the file {\tt toprex.info} 
will be assigned to the Logical Unit Number (LUN) 6.

As described above, all parameters can also be specified in two separate
routines ({\tt main.f} or {\tt TXPYIN} and {\tt TopReX}).
In addition, the user can use a special file {\tt run.dat}
to read in input parameters (see appendix~3 for details).

%%%%%%%%%%%%%%%%%%%%%%%%%%%%%%%%%%%%%%%%%%%%%%%%%%%%%%%%%%%%%%%%%%%%%%%%%%%
\section{\bf \TR linkage to \PYT}

Along with the \TR package comes a GNUmakefile which links the necessary 
libraries and the source code together and produces an executable.
The program package was tested under Solaris, Digital~Unix, and Linux and 
should also run on 
other platforms. The libraries \PYT and {\tt cernlib} are assumed to be 
accessible under the path {\tt /cern/pro/lib}. 
If this is not the case the user has to change the library search path 
accordingly. The TAUOLA library is provided as source code and
 will be 
compiled as separate library. The \TR source code, the slightly modified 
\PYT subroutines {\tt PYDECY}, {\tt PYEVNT}, {\tt PYSHOW}, and
{\tt PYUPIN} (for \PYT version 6.136 and 
newer) are compiled together  with the example main program listed below. To
 take into account the different cases of the different versions of \PYT, 
two targets which produce executables are provided in the GNUmakefile. For 
\PYT before version 6.157 the executable has to be produced by the
 command  {\tt gmake toprex.exe} while {\tt gmake TX.exe} has to be used 
otherwise.
 If the user is already familiar with the event generator \PYT, the \TR 
package may easily 
be integrated in existing program structures (user main program, 
analysis routines and detector simulation).

The user has to keep in mind to compile and link all routines of the \TR
package beside his routines.
To achieve this, either the user program file names have to be added in the
provided GNUmakefile, e.g. they may be appended to the {\tt USEROBJ} variable 
or all
\TR related file and library names have to be included in the user makefile.

%%%%%%%%%%%%%%%%%%%%%%%%%%%%%%%%%%%%%%%%%%%%%%%%%%%%%%%%%%%
\section{\bf Interface and program flow}

Global \TR parameters can be accessed and changed via the common block 
{\tt TXPAR}:   % which is of the form
\bigskip

\noindent
\drawbox{\label{txpar}\tt COMMON/TXPAR/ Ipar(200), Rpar(200)} .
\vspace*{-2ex}
\newline
Integer values like the \TR process number to be chosen by the user are 
stored in the array {\tt Ipar} as indicated in Table~\ref{txnump}.
Floating point values like the hadronic centre-of-mass energy available for 
collisions are stored in the array {\tt Rpar}.
A complete list of the parameters and their default values is given in 
appendix~4. 
As a minimal requirement the number of a subprocess has to be specified. 
Further parameters like the hadronic centre-of-mass energy, particle masses 
etc. are taken from \PYT. Two examples how to use \TR in conjunction with 
\PYT are given in appendices~1 and~2.

In analogy to the \PYT common block {\tt /PYDAT3/}, which offers the 
possibility to switch {\sf on/off} specific decay channels of particles after 
the generation of the hard subprocess, the \TR common block {\tt /TXRDEC/}
 provides this possibility for $W^{\pm}$, $H^{\pm}$, and $t$-quark 
decays which are part of the hard subprocess.
For convenience, the numbering scheme is adopted to that of PYTHIA.  
Numbers of the {\it Individual Decay Channels} (IDC's) for $W$ and
$Z$-bosons are 
identical to those of \PYT (see \PYT Manual~\cite{pythia}). IDC numbers for 
$H^{\pm}$ are 
identical to those of the $W^{\pm}$ boson with one additional channel 
(IDC=21): 
$H^{\pm} \to \bar b t^* \to \bar b W b$. 
For top quark decays the IDC's are defined as follows:
\begin{itemize} 
\item[1.] (IDC = 1) SM decay, $t \to bW$
\item[2.] (IDC = 2) $t \to bW$ decay via SM and beyond SM interactions
\item[3.] (IDC = 3) $t \to b\,H^{\pm}$ 
\item[4.] (IDC = 4,..,11) correspond to top quark decay via Flavour Changing 
  Neutral Current interactions
\end{itemize} 
%\vspace*{-1ex}
The common block {\TTF /TXRDEC/} is defined as follows: 
\bigskip

\noindent
\drawbox{COMMON/TXRDEC/ MID(4,0:30), BRF(4,30), FID(4,30,5), BRS(4,5)}%
\vspace*{-2ex}
\newline
with the integer parameter array {\tt MID} and the double precision parameter 
arrays {\tt BRF, FID and BRS}, explained below.

\begin{entry}
\itemc{Purpose:} to access particle decay data and parameters.
The first index I (I=1,2,3,4) of the arrays corresponds to $t$-quark, $W$, $H$,
 and $Z$-boson decays, respectively.
\iteme{MID(I,IDC) :} {\sf on/off} switch for IDC (meaning identical to \PYT 
parameter {\TTF MDME })
\iteme{MID(I, 0)  :} total number of allowed IDC
\begin{subentry}
\iteme{ = -1} coupling switched {\sf off } 
\iteme{ = 0 } channel is switched {\sf off } for decay, but it contributes
to the total width
\iteme{ = 1 } channel is switched {\sf on} 
\iteme{ = 2 } channel is switched {\sf on} for {\it particle } 
             but {\sf off } for {\it anti-particle } 
\iteme{ = 3 } channel is switched {\sf on} for {\it anti-particle } 
             but {\sf off } for {\it particle } 
\iteme{ = 4,5 } are not used in current version 3.25 of \TR.
\end{subentry}
\iteme{BRF(I,IDC)  :} branching fraction for given IDC
\iteme{FID(I,IDC,j):} contains the KF code for decay products of a given IDC, 
  \newline
\phantom{FID(} meaning identical to \PYT parameter {\TTF KFDP} 
\iteme{BRS(I,IDC)  :} sum of branching fractions for different decay
             groups 
\end{entry}
\medskip

%The schematic structure of \TR and its data flow is shown in figure 
%\ref{dataflow} with an example of CMS detector simulation interfaced to \PYT.
%The subroutines {\tt TopReX\_Input} and {\tt TopReX\_NT} are optional 
%depending on the used mode.

%%%%%%%%%%%%%%%%%%%%%%%%%%%%%%%%%%%%%%%%%%%%%%%%%%%%%%%%%%%%%%%%%%%%%%%%%%%
\section{\bf Description of hard processes} 
Some comments on the hard processes of \TR, listed in table~\ref{txnump}, are 
made in the following. 
The kinematics of a hard process can be specified
by setting the values of the \PYT parameters  {\TTF CKIN(1:4)} in the
{\TTF main.f} or {\TTF TXPYIN} routine. The values of the {\tt CKIN} 
parameters are copied
to the \TR parameters {\TTF Rpar(101:104)} in the subroutine {\TTF TOPREX}.
Three processes ({\tt Ipar(1) = 1, 7, 10}) are singular at 
$\hat p_{\perp} \, \to \, 0$. To avoid the singularities a conventional
kinematical cut on ($\hat p_{\perp}$) has to be used (a minimal value
of $20$~GeV is recommended):

{\TTF CKIN(3) = 20.d0} (copied during initialization to the \TR parameter 
{\TTF Rpar(103)})  

The $W$-gluon fusion processes ({\TTF Ipar(1) = 1,2}),
$W* \to \tau \nu$ ({\TTF Ipar(1) = 6,7}) and 
$H^{\pm}* \to \tau \nu$ ({\TTF Ipar(1) = 9,10}) 
need some comments concerning the problem of double counting, given below.
Following the work of \cite{boos} a simple method for the generation of such 
reactions is applied  (see also \cite{srstau}). 

\noindent {$\diamond$} {\bf Single top Production:} $W$-gluon fusion,
{\tt Ipar(1) = 1, 2} \\
Two different subprocesses can be used for the generation of single top
production in $W$-gluon fusion processes: $2 \to 2$ ({\tt Ipar(1) = 1}) with
{\TTF CKIN(3) = 0.0} and $2 \to 3$ ({\tt Ipar(1) = 2}) with
{\TTF CKIN(3) = 20.0}. The additional $\bar b$-quark in the first kind of 
events
appears in the initial state and can be found in the {\TTF PYJETS} list.
Such an event will be accepted
for further analysis, if the transverse momentum of this additional 
$\bar b$-quark does not exceed some threshold $p_{\perp 0}(b)$ (typically of 
the order $\sim 10$~GeV).
An event of the second kind ($q g \to q' t \bar b$ process)
 will be accepted if the transverse momentum of the final $\bar b$-quark from
 the hard process will be greater than the $p_{\perp 0}(b)$ value.

\noindent {$\diamond$} {{\bf $W^* \to \tau \nu$} ({\tt Ipar(1) = 6,7}) and 
{\bf $H^{\pm *} \to \tau \nu$} ({\tt Ipar(1) = 9,10}) processes}{\bf :} \\
These processes are generated in the same way as for the
single top production process through the $W$-gluon fusion mechanism.
However, the user should examine the transverse
momentum of the intermediate $W^*$ or $H^*$-boson, here. 
For $W^*$ production processes it is recommended to choose threshold values 
above 
\[ p_{\perp 0}(W) \approx 40 \,\,\, {\rm GeV}
\]
For the charged Higgs production the $p_{\perp 0}(H)$ value depends on the 
mass of the charged Higgs and should be figured out by the user.

For processes with a charged Higgs ($H^{\pm}$) the default 
values for $M(H^{\pm})$ and $\tan\beta$ are given below:
\begin{center} 
\begin{tabular}{ll} 
 $M(H^{\pm}) = 300$~GeV & {\TTF Rpar(51) } \\
 $\tan\beta  = 50$      & {\TTF Rpar(55) } \\
\end{tabular} 
\end{center} 
To change these values the corresponding parameters {\TTF Rpar(51),
 Rpar(55)} have to be modified in the subroutine {\TTF TOPREX}.

\noindent {$\diamond$} {\bf Top anti-top quark pair production:} $t\bar{t}\rightarrow
 bW^+\bar{b}W^-\rightarrow 6~\mbox{fermions}$ \\ 
The processes {\tt Ipar(1)=20,21,22} provide the $t\bar{t}$ production in the
Breit-Wigner approach with off-shell $W$ bosons \cite{Kleiss88}
(see \cite{Beneke99} for details about the top quark production).
The $t\bar{t}$ processes ({\tt Ipar(1)=50,
51,52}) in the on-shell 
approximation 
\cite{Flesch98} provide the possibility to switch on an 
intermediate neutral Higgs boson which becomes resonant in the case of a 
Higgs mass above the $t\bar{t}$ production threshold. 
Below, the $t\bar{t}$ production via a Higgs resonance is switched off. 
The matrix elements calculate the coherent contribution of the resonant Higgs 
boson and the non resonant $t\bar{t}$ background. 
They are also valid in supersymmetric extensions of the standard model in case
of squark masses above 400~GeV, 
since the contributions of such heavy squarks are negligible \cite{Daw96}. 
The scalar and pseudoscalar Yukawa couplings to top quarks are described 
by the parameters {\tt Rpar(81)$=a$} and {\tt Rpar(82)$=\tilde{a}$} 
respectively. The standard model couplings of the Higgs boson to the $W$ and 
$Z$ vector bosons are realized with a multiplicative factor, the relative 
coupling strength 
{\tt Rpar(80)$=g_{VV}$}, which depends as the reduced Yukawa couplings in a 
two Higgs doublet model 
on the ratio of the vacuum expectation values of the Higgs doublets. 
In the $t\bar{t}$ matrix elements used here this factor is important, 
to take into account the total decay width of the Higgs boson, which is also 
printed as control output after initialization of the processes.  
The user has to provide the values for the Higgs boson couplings. Several 
examples are listed in Table~\ref{higgscouplings}.

\begin{table}[ht] 
\begin{center} 
\begin{tabular}{lccc}
                             & {\tt Rpar(80)} &{\tt Rpar(81)} & {\tt Rpar(82)}
                                                                   \\[0.5ex]  
 Standard model Higgs boson: & $g_{VV}=1$, & $a=1$,     & $\tilde{a}=0$ \\
 Pseudoscalar Higgs boson:   & $g_{VV}=0$, & $a=0$,     & $\tilde{a}=1$ \\
 CP violating Higgs boson:   &             & $a\neq 0$, & $\tilde{a}\neq 0$
\end{tabular} 
\end{center}
\vspace*{-3ex}
\caption{ 
  \label{higgscouplings}
  Examples of the Higgs boson couplings {\tt Rpar(80)$=g_{VV}$}, 
{\tt Rpar(81)$=a$} and {\tt Rpar(82)$=\tilde{a}$} 
  of the $t\bar{t}$ production processes {\tt Ipar(1)=50,51,52}.
}
\end{table} 

%In the next section a simple example shows how to run {{\sf\bf TopReX}}.

\section{\bf Comparison of parton distributions}

To demonstrate the reliability of the processes implemented in \TR, they are 
compared to the default processes of \PYT 6.1. for $pp$ collisions at 
$\sqrt{s}=14~\mbox{TeV}$ (LHC).

\begin{figure}[t]
  \unitlength 1cm
  \begin{picture}
(11.5,20.0)
    \put(2.2,10.5){\scalebox{0.6}{\includegraphics{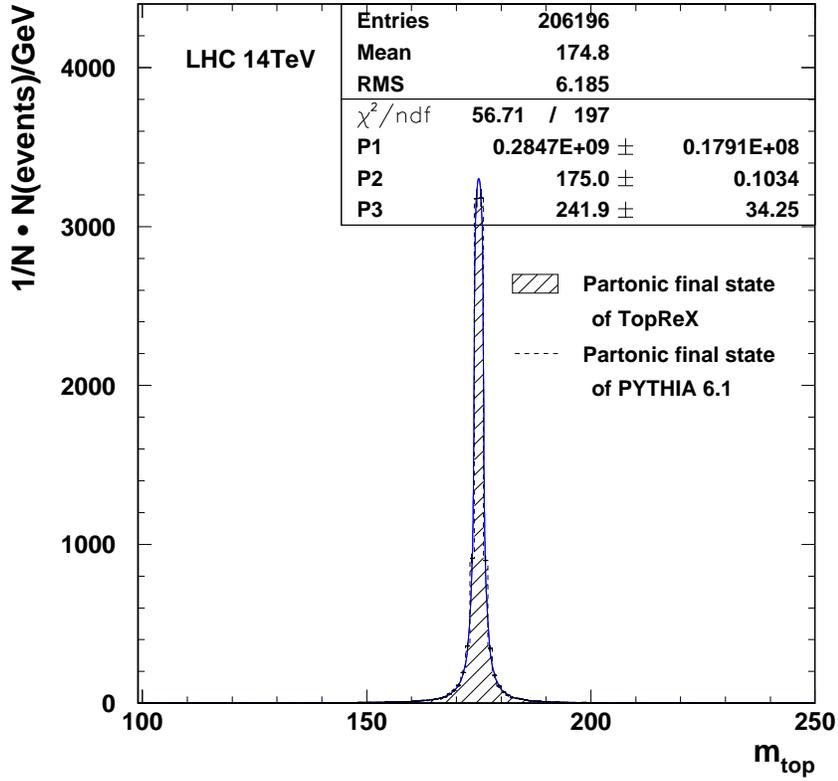}}}
    \put(2.2,-0.6){\scalebox{0.6}{\includegraphics{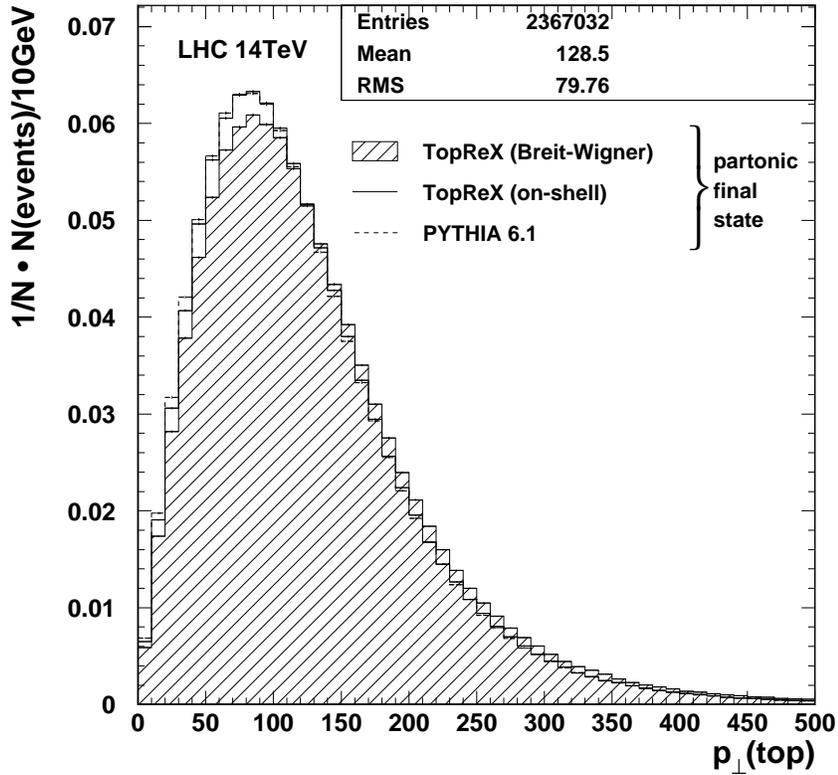}}}
  \end{picture}
  \begin{center}
    \parbox{15cm}{
      \caption{ The top quark mass (upper plot). The distribution of the 
       $2\rightarrow 6$ $t\bar{t}$ matrix elements of \TR in the Breit-Wigner 
       approach shows agreement 
               with the prediction of the $2\rightarrow 2$ $t\bar{t}$ 
        matrix elements of \PYT 6.1. The natural width of the top quark mass 
        is fitted with a Breit-Wigner function. 
        The lower plot shows the $p_{\perp}$ spectrum of the top quarks.
       The prediction of \PYT 6.1 coincides with the on-shell approach of \TR. 
        The Breit-Wigner approach, used here with a completely different $Q^2$
        scale,   gives quite similar results. 
        \label{tmass} } }
  \end{center}
\end{figure}

In the upper plot of figure \ref{tmass} the distribution of the top quark mass
 is shown for the two event generators. The $2\rightarrow 6$ $t\bar{t}$ matrix 
elements  in the Breit-Wigner approach are in good agreement with the 
$2\rightarrow 2$ $t\bar{t}$ matrix elements of \PYT 6.1. The lower plot shows 
that the $p_{\perp}$ spectrum of the top quarks are in agreement for the 
different event matrix elements of the compared generators.

\begin{figure}[t]
  \unitlength 1cm
  \begin{picture}
(11.5,21.0)
    \put(2.2,10.5){\scalebox{0.6}{\includegraphics{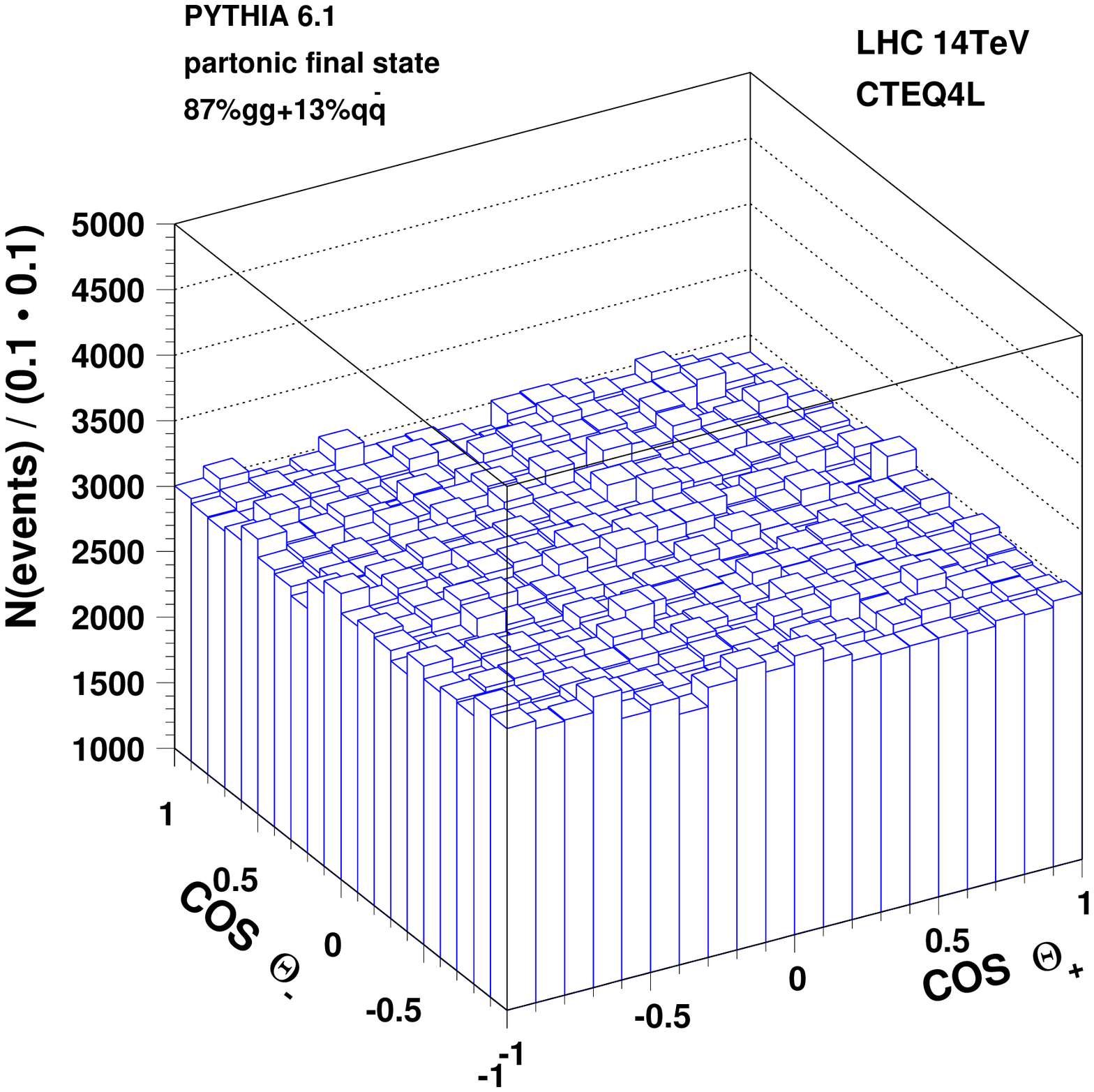}}}
    \put(2.2,-0.8){\scalebox{0.6}{\includegraphics{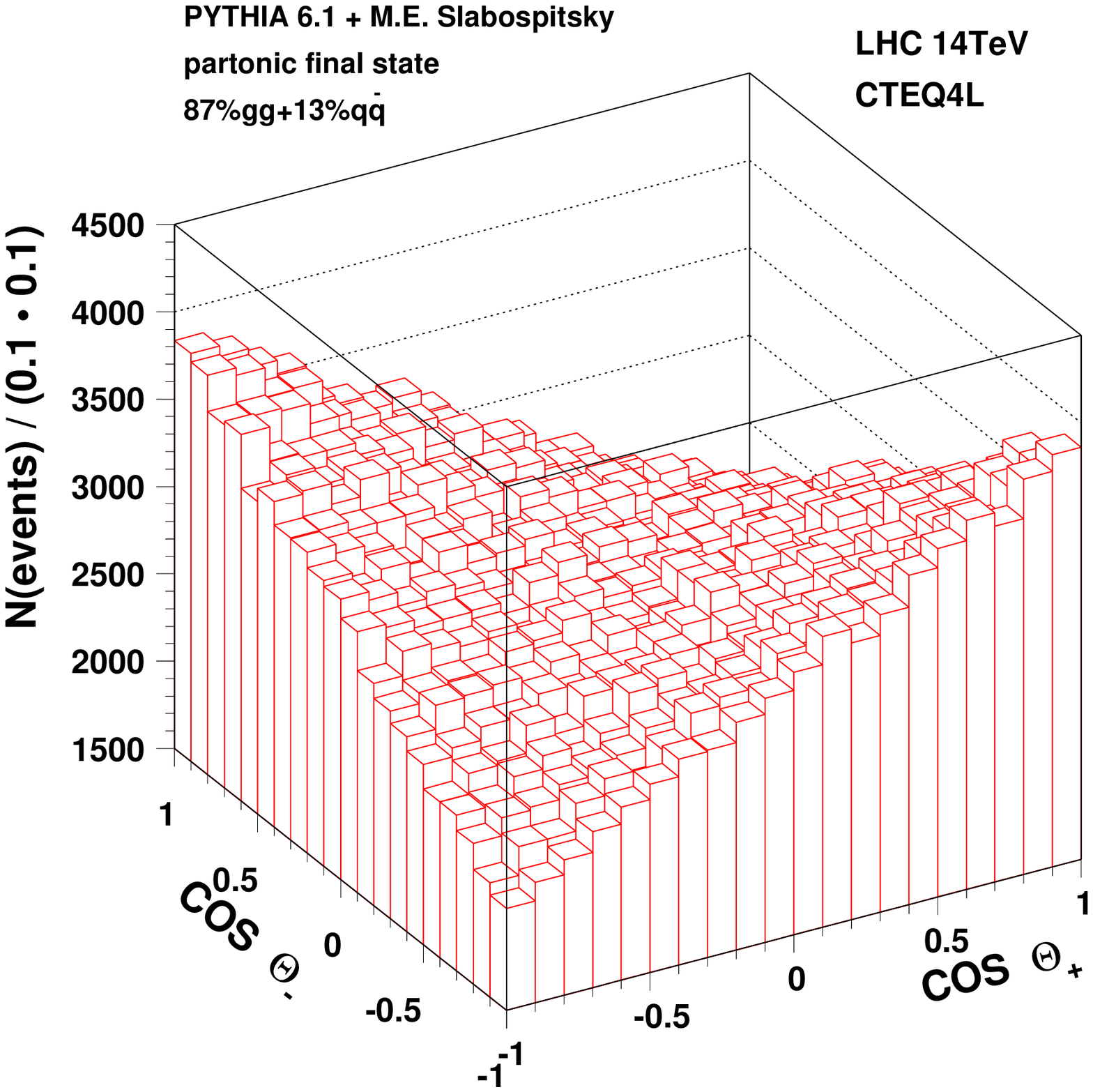}}}
  \end{picture}
  \begin{center}
    \parbox{15cm}{
      \caption{ $t\bar{t}$ spin correlation in the helicity basis.
        The prediction of the $2\rightarrow 2$ $t\bar{t}$ matrix elements of 
        \PYT 6.1 show no correlation between the two leptons (upper plot). 
        In contrast, the standard model predicts the correlation shown in the 
        lower plot.
        This result is obtained by the $2\rightarrow 6$ $t\bar{t}$ matrix 
        elements implemented in \TR.
        In the Breit-Wigner approach (shown here) the asymmetry coefficient 
        reads ${\cal{A}}_{\mbox{\scriptsize SM}}=0.32$.
        In the on-shell approach the coefficient amounts to 
        ${\cal{A}}_{\mbox{\scriptsize SM}}=0.33$.
        \label{ttbarspincorr2comp} } }
  \end{center}
\end{figure}

In figure \ref{ttbarspincorr2comp} the spin correlation between the top quark
 decay products is shown for the $t\bar{t}$ production in the di-leptonic 
decay channel. The angles of the two leptons are evaluated in the helicity 
basis  convenient at the LHC \cite{Alt00}, \cite{Ber01}. While the 
$2\rightarrow 2$ matrix elements of \PYT cannot include the spin correlation, 
the $2\rightarrow 6$ matrix elements of \TR do. Conclusively, the asymmetry 
coefficient of the spin correlation obtained using default \PYT gives 
vanishing values. In contrast the standard model prediction, reproduced by 
\TR to leading order, yields the asymmetry coefficient ${\cal{A}}_{SM}=0.32$ 
using the parton densities of CTEQ4L.

\begin{figure}[t]
  \unitlength 1cm
  \begin{picture}
(11.5,21.0)
    \put(0.5,14.7){\scalebox{0.4}{\includegraphics{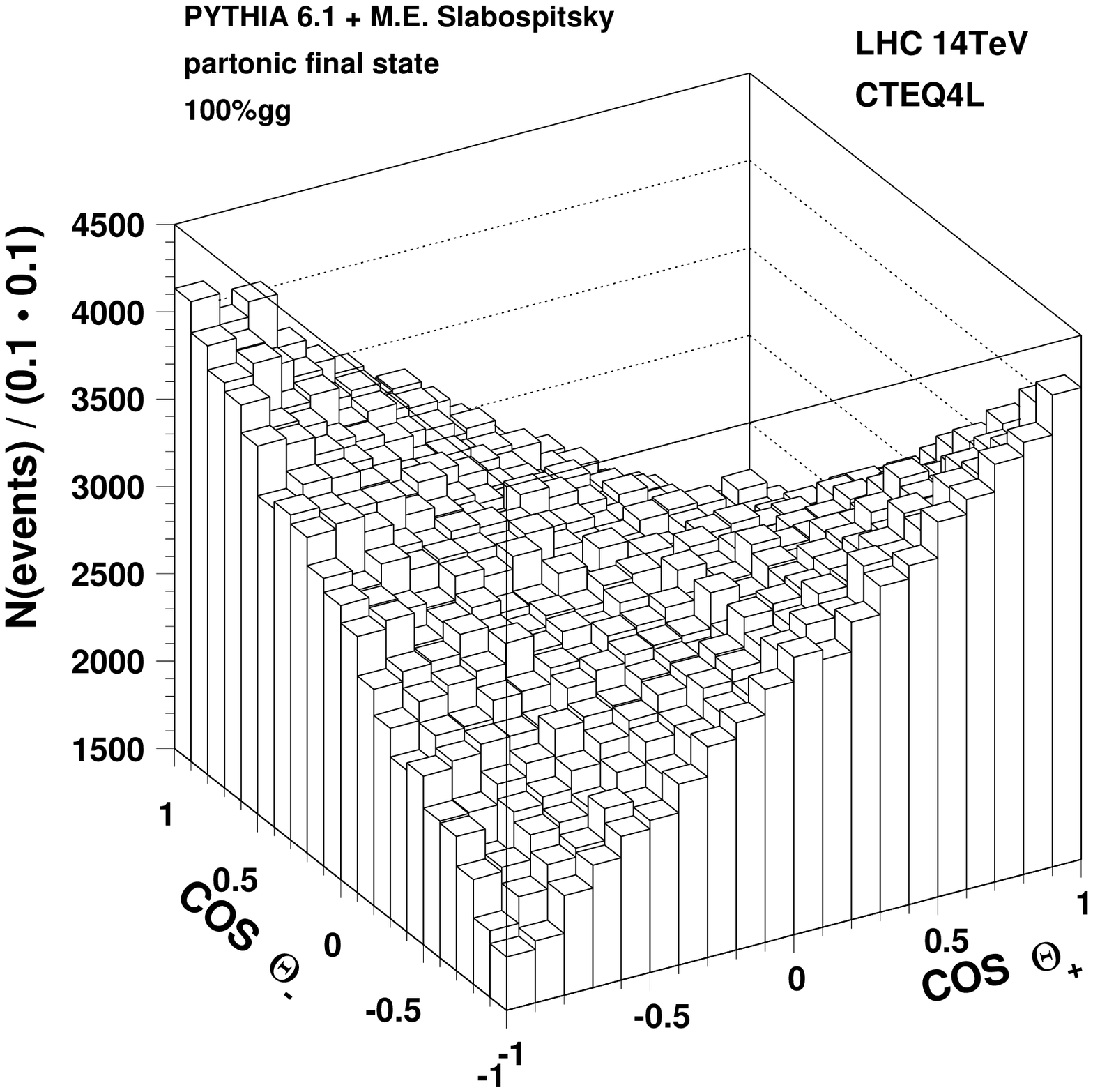}}}
    \put(2.6,20.5){${\cal{A}}_{\mbox{\scriptsize}}=0.43$}
    \put(0.5,7.1){\scalebox{0.4}{\includegraphics{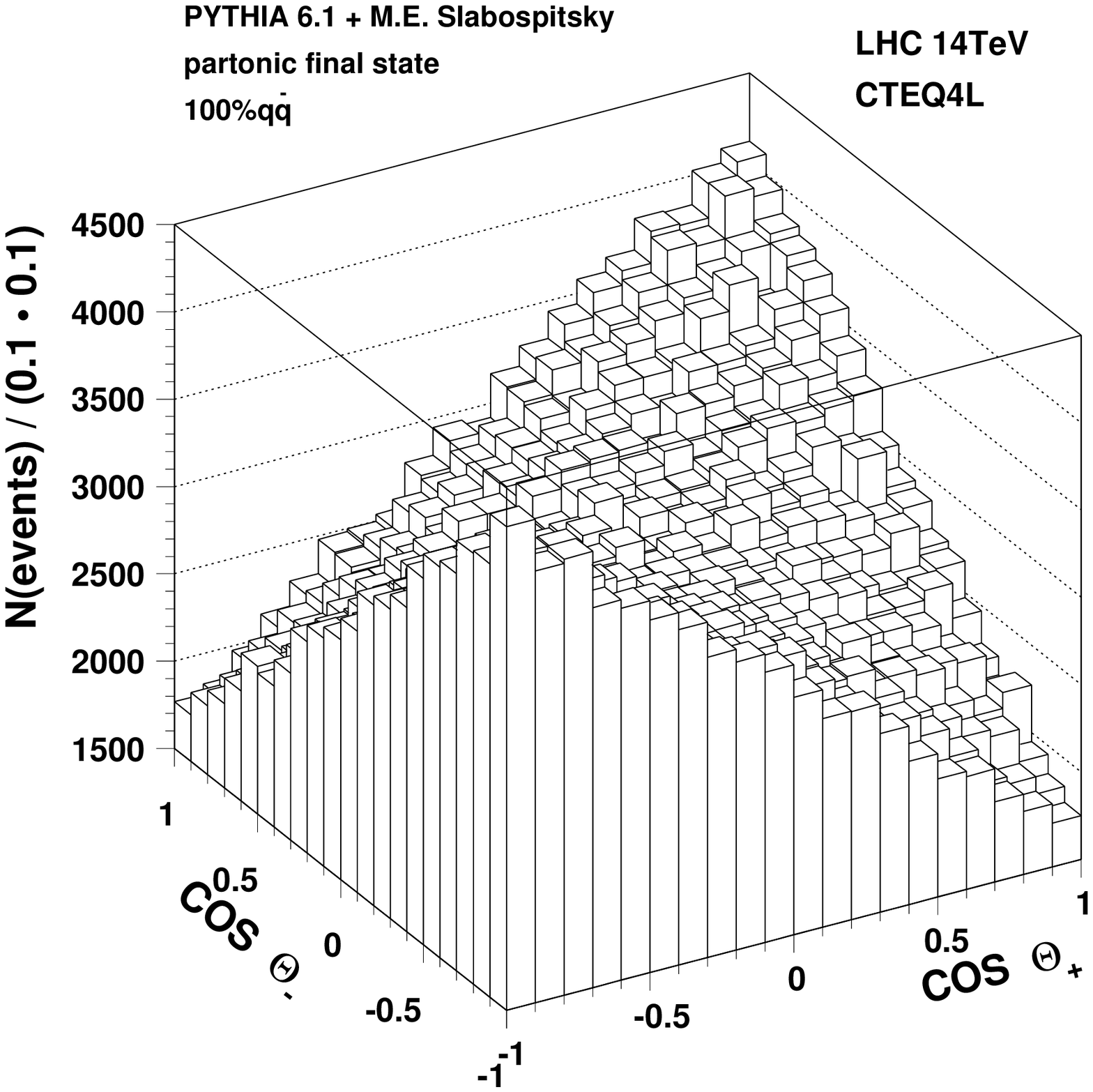}}}
    \put(2.3,12.9){${\cal{A}}_{\mbox{\scriptsize}}=-0.46$}
    \put(0.5,-0.5){\scalebox{0.4}{\includegraphics{trxcorr.eps}}}
    \put(2.6,5.3){${\cal{A}}_{\mbox{\scriptsize SM}}=0.32$}
    \put(8.3,14.7){\scalebox{0.4}{\includegraphics{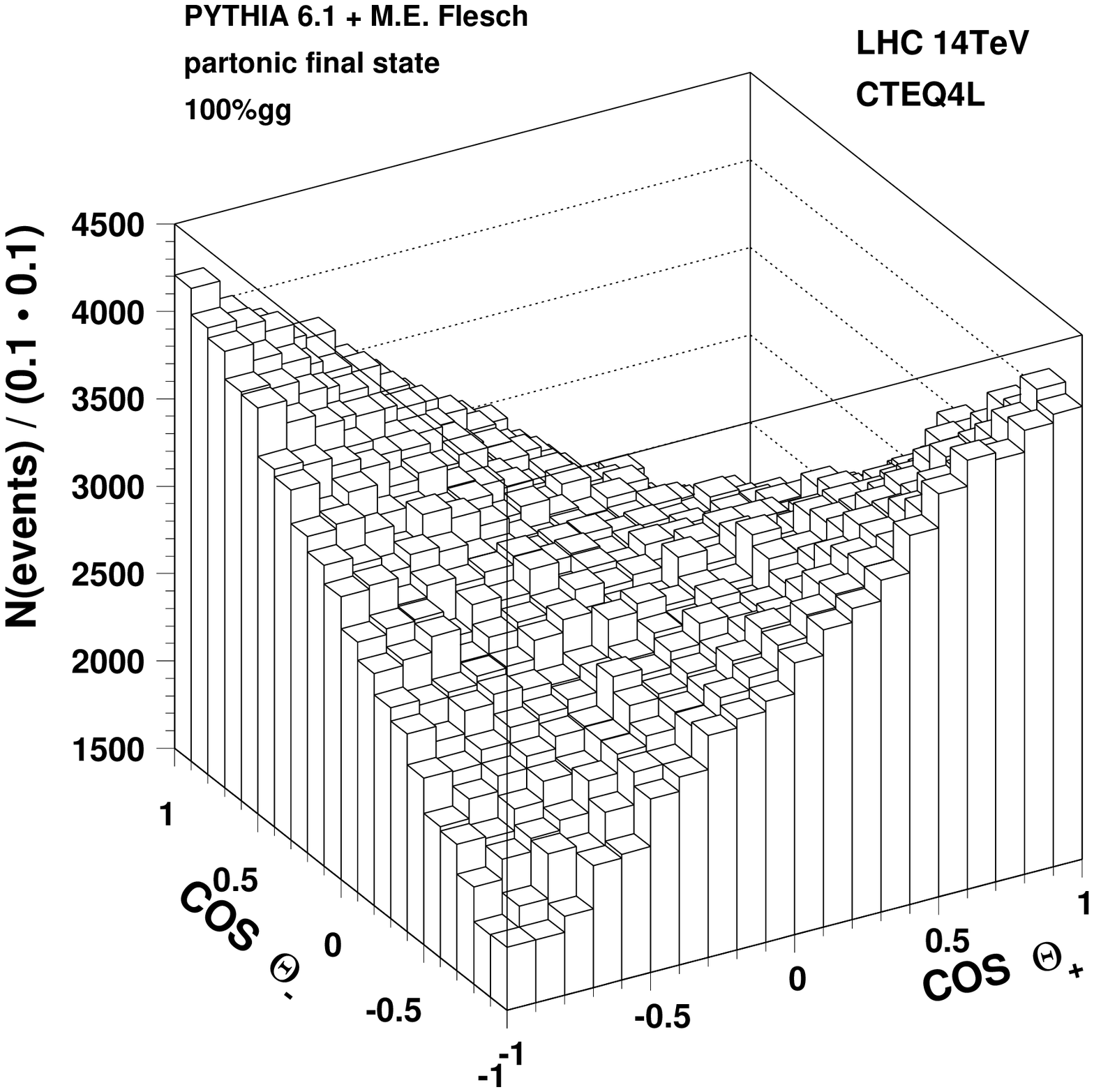}}}
    \put(10.4,20.5){${\cal{A}}_{\mbox{\scriptsize}}=0.45$}
    \put(8.3,7.1){\scalebox{0.4}{\includegraphics{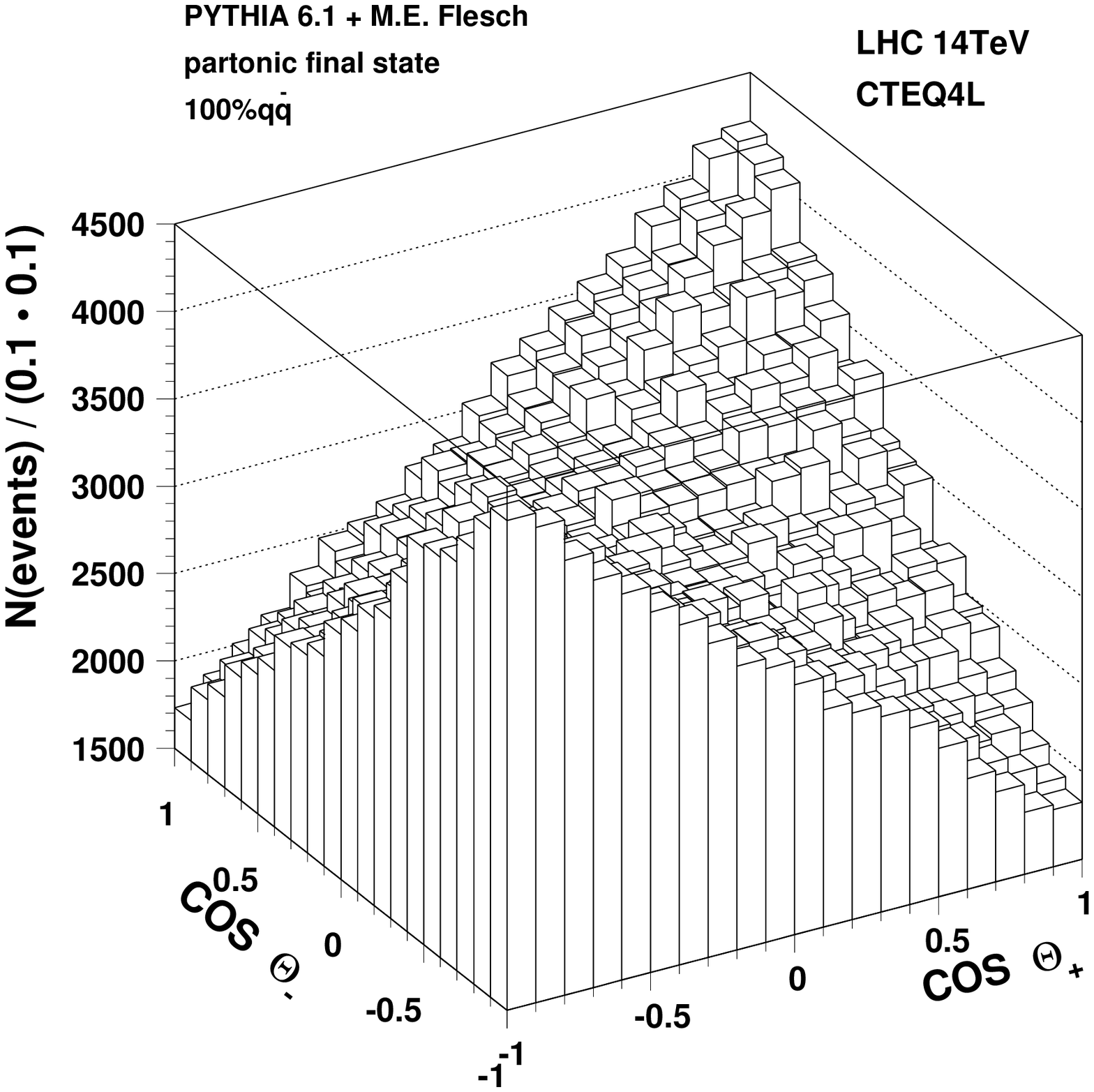}}}
    \put(10.1,12.9){${\cal{A}}_{\mbox{\scriptsize}}=-0.46$}
    \put(8.3,-0.5){\scalebox{0.4}{\includegraphics{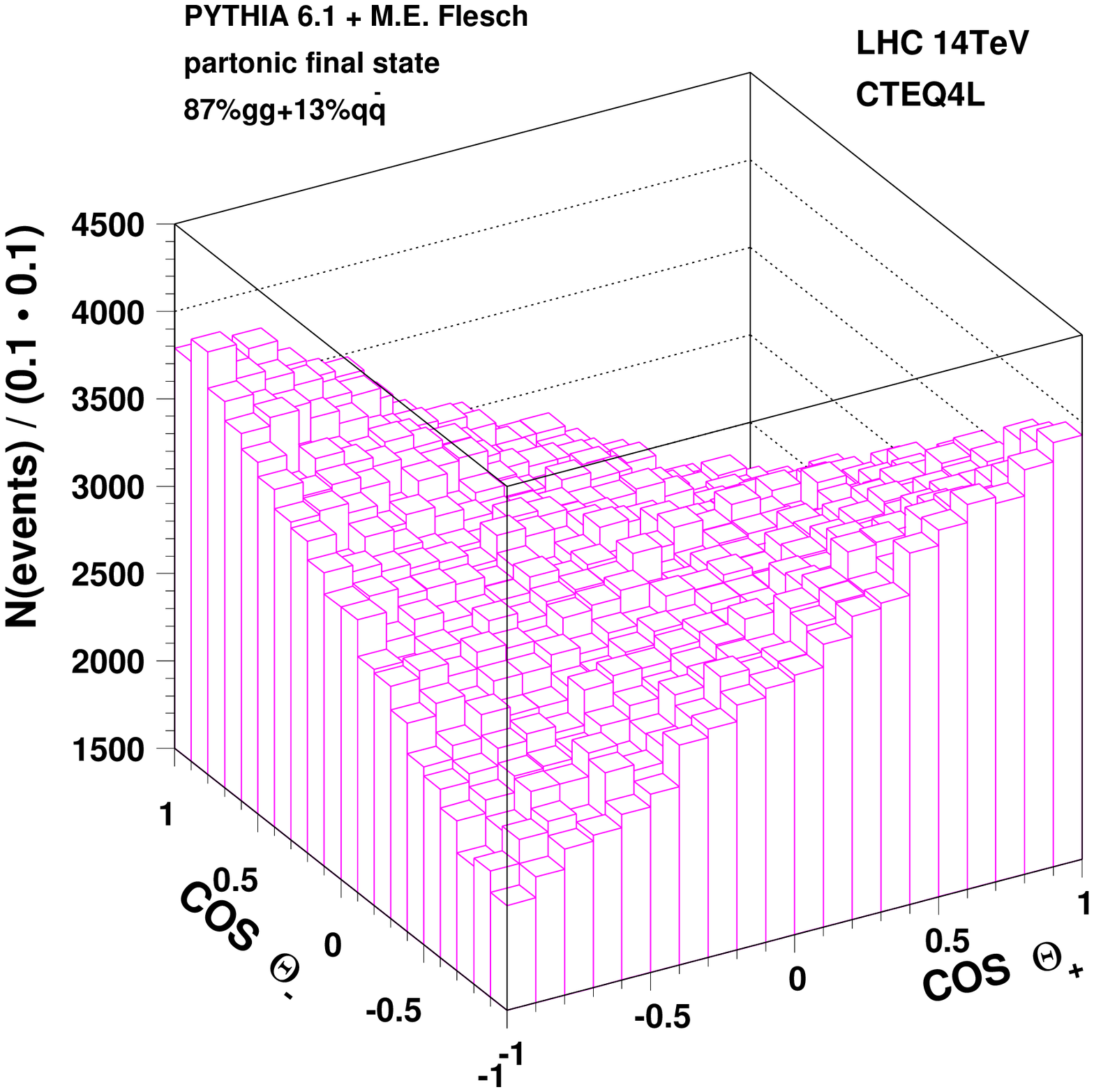}}}
    \put(10.4,5.3){${\cal{A}}_{\mbox{\scriptsize SM}}=0.33$}
  \end{picture}
  \begin{center}
    \parbox{15cm}{
      \caption{ $t\bar{t}$ spin correlation of \TR matrix elements in the 
        helicity basis. To the left the prediction of the $2\rightarrow 6$ 
        $t\bar{t}$ matrix elements in the Breit-Wigner approach is shown.
        To the right, the same is shown for the $2\rightarrow 6$ $t\bar{t}$ 
        matrix elements in the on-shell approach.
        At the top only the contribution of the gluon gluon fusion is plotted.
      In the center only the contribution of the quark anti-quark annihilation
        is given and at the bottom both production processes
        are taken into account. The corresponding asymmetry coefficients are 
        indicated in the plots. 
        The Breit-Wigner and on-shell approaches are in very good agreement. 
        \label{ttbarspincorr6comp} } }
  \end{center}
\end{figure}

Figure \ref{ttbarspincorr6comp} shows a comparison of the $t\bar{t}$ spin 
correlation with help of the helicity basis 
for the matrix elements in the Breit-Wigner (left) 
and on-shell (right) approach, implemented in \TR. The contribution to this 
correlation of the gluon gluon fusion (top) and 
the quark anti-quark annihilation processes (center) is given separately. 
The predictions of the matrix elements in the Breit-Wigner approach and in the
 on-shell approach agree quite well.

%%%%%%%%%%%%%%%%%%%%%%%%%%%%%%%%%%%%%%%%%%%%%%%%%%%%%%%%%%%%%%%%%%%%%%%%%%%%%
\section{\bf Conclusions}

The specialised event generator \TR, 
which provides the simulation of several heavy 
quark production processes in hadronic collisions has been described. 
As cross check many comparisons of parton distributions 
between the matrix elements implemented in \PYT and \TR have been applied.
The agreement is very good except in cases like the spin correlation where 
default \PYT is not able to reproduce the standard model predictions.

%%%%%%%%%%%%%%%%%%%%%%%%%%%%%%%%%%%%%%%%%%%%%%%%%%%%%%%%%%%%%%%%%%%%%%%%%%%%%
\section* {\bf Acknowledgements}

Great thanks to  
S.~Abdullin,     W.~Bernreuther, E.~Boos,      K.~Datsko,    D.~Denegri, 
V.~Drollinger,   G.~Fl\"ugge,    U.~Heintz,    V.~Ilyin,     R.~Mehdiyev, 
A.~Miagkov,      D.~Murashev,    A.~Nikitenko, V.~Obraztsov, R.~Schulte,  
S.~Stepanov, and O.~Yushchenko  
for fruitful discussions and support during the development of \TR and 
this manual.

\newpage

%%%%%%%%%%%%%%%%%%%%%%%%%%%%%%%%%%%%%%%%%%%%%%%%%%%%%%%%%%%%%%%%%%%%%%%%%%%%%

\newpage
%%%%%%%%%%%%%%%%%%%%%%%%%%%%%%%%%%%%%%%%%%%%%%%%%%%%%%%%%%%%%%%%%%%%%%%%%%%%%%

\section*{\bf Appendix~1.  }
\label{app1}

The example main program shows the usage of a particular (\PYT external) \TR 
subprocess.  \PYT parameters like particle
 masses, etc. and the \TR process number
({\tt IPROC} parameter) are specified in the {\tt MAIN\_1} routine shown below.
To access any \TR subprocess the parameter {\tt IPROC} 
has to be set to the desired subprocess (here {\tt IPROC = 50} 
specifies the $t\bar{t}$ production in the on-shell approach).
The choice has to be made before the call of the {\tt TOPREX} subroutine.
The further initialisation of \PYT is done by a call to the subroutine 
{\tt PYINIT}.

\small
\begin{verbatim}
      PROGRAM MAIN_1 

      IMPLICIT NONE
      CHARACTER*8 FRAME, BEAM, TARGET
      DOUBLE PRECISION ECM     ! CMS energy
      INTEGER IPROC            ! number of TopReX process
      INTEGER I

      EXTERNAL PYDATA
      INTEGER MSTU, MSTJ,KCHG, MDCY,MDME, KFDP, MSEL, MSELPD, MSUB
      INTEGER KFIN, MSTP, MSTI
      DOUBLE PRECISION PARU, PARJ, PMAS, PARF, VCKM, BRAT, CKIN
      DOUBLE PRECISION PARP, PARI
      COMMON /PYDAT1/ MSTU(200),PARU(200),MSTJ(200),PARJ(200)
      COMMON /PYDAT2/ KCHG(500,4),PMAS(500,4),PARF(2000),VCKM(4,4)
      COMMON /PYDAT3/ MDCY(500,3),MDME(4000,2),BRAT(4000),KFDP(4000,5)
      COMMON /PYSUBS/ MSEL,MSELPD,MSUB(500),KFIN(2,-40:40),CKIN(200)
      COMMON /PYPARS/ MSTP(200),PARP(200),MSTI(200),PARI(200)

      FRAME  = 'CMS'
      BEAM   = 'P'
      TARGET = 'P'
      ECM    = 14000.d0   ! LHC cms energy (in GeV)

      IPROC = 50          ! tt~ production

      PMAS(6,1)  = 175.d0     ! top mass
      PMAS(5,1)  =   4.8d0    ! bottom mass
      PMAS(23,1) =  91.187d0  ! Z0 mass
      PMAS(24,1) =  80.41d0   ! W mass
      PMAS(25,1) = 400.d0     ! Higgs mass

      MSTP(51) = 7           ! PDF: CTEQ5L

      CALL TOPREX(FRAME, BEAM, TARGET, IPROC, ECM)

      CALL PYINIT(FRAME,BEAM,TARGET, ECM)

      DO I = 1,10 
        PRINT*,'Event number I=',I
        CALL PYEVNT()
        CALL PYHEPC(1)
        IF (I.LE.2)  CALL PYLIST(2)

c...    at this place analysis routines and/or
c...    detector simulations may be called

      ENDDO

      END
\end{verbatim} 
\normalsize

\newpage
%%%%%%%%%%%%%%%%%%%%%%%%%%%%%%%%%%%%%%%%%%%%%%%%%%%%%%%%%%%%%%%%%%%%%%%%%%%%%%

\section*{\bf Appendix~2.  }
\label{app2}
Contrary to the previously described {\tt MAIN\_1} routine, this example 
considers the possibility to set all \PYT parameters in a special {\tt TXPYIN} 
subroutine. In addition, a few parameters are read in by the 
{\tt TXRINT} subroutine from an external input data file, named {\tt run.dat} 
(see appendix~3 for details).
An example of a special routine for user's analysis 
({\tt USRPRO}) is described, too.

\small
\begin{verbatim}
      PROGRAM MAIN_2 

      IMPLICIT NONE 
      DOUBLE PRECISION Ecm        ! CMS energy
      INTEGER IPROC               ! no. of Toprex process
      INTEGER Ntot, N, MODE, IER  ! internal variables
*...............................................................
      Ecm  = 14000.d0          ! CMS energy (in GeV) for LHC option 

      call TXWRIT(-6)          ! LUN=6, open file with output information    
* 
      call TXRINT('iproc=', iproc, ier) ! no. of TopRex process 
      call TXRINT('Ntot=',   Ntot, ier) ! number of events to be generated
* 
      call TXPYIN(IPROC, Ecm)  ! all PYTHIA will be specified there   
* for compilation user's program WITHOUT TopRex package comment next line 
      if(Iproc.ge.1) call TOPREX('CMS','P','P', IPROC, Ecm) ! TopRex init.
* . . . . . . . . 
*
      call PYINIT('CMS','P','P', Ecm)
      call PYSTAT(4)
c... call analysis/detector simulation  routine for initialisation
      mode = -1
      call USRPRO(mode, Ntot) 
*... end of initialisation
      DO N = 1, Ntot 
        if(mod(N,1000).eq.1) write(*,*)' event no=', N 
        call PYEVNT()
        call PYHEPC(1)
c... call analysis/detector simulation routine for running
        mode = 0
        call USRPRO(mode, N) 
      ENDDO
*
      call PYSTAT(1)      ! Brief statistics output from PYTHIA
*... call analysis/detector simulation routine for closing
       mode = 1
       call USRPRO(mode, N) 
*...
      STOP
      END
\end{verbatim} 

\noindent The following example routine {\tt TXPYIN } is used to set 
\PYT parameters.

\begin{verbatim} 
      SUBROUTINE TXPYIN(IPROC, Ecm)       

      IMPLICIT NONE

*...Standard PYTHIA ( v. >= 6.1) commons for initialization.
      EXTERNAL PYDATA
      INTEGER MSTU, MSTJ, KCHG, MDCY, MDME, KFDP, MSEL, MSUB, KFIN,
     &        MSTP, MSTI, MSELPD
      DOUBLE PRECISION
     &  PARU, PARJ, PMAS, PARF, VCKM, BRAT, CKIN, PARP, PARI
      COMMON /PYDAT1/ MSTU(200),PARU(200),MSTJ(200),PARJ(200)
      COMMON /PYDAT2/ KCHG(500,4),PMAS(500,4),PARF(2000),VCKM(4,4)
      COMMON /PYDAT3/ MDCY(500,3),MDME(4000,2),BRAT(4000),KFDP(4000,5)
      COMMON /PYSUBS/ MSEL,MSELPD,MSUB(500),KFIN(2,-40:40),CKIN(200)
      COMMON /PYPARS/ MSTP(200),PARP(200),MSTI(200),PARI(200)
*.......................
      INTEGER IPROC, I   
*............................................
*
      MSTP(51) = 8            !  PDF CTEQ5M1 (PYTHIA6.1..)  
*.................masses........
      PMAS(6,1)  = 175.d0     ! top mass
      PMAS(5,1)  =   4.8d0    ! bottom mass
      PMAS(24,1) =  80.41d0   ! W mass
      PMAS(25,1) = 400.d0     ! Higgs mass
*...KINEMATICS
*
      CKIN(3) = 20.d0    ! PT HAT LOW CUT
      CKIN(4) = 200.d0   ! PT HAT UP  CUT
*..........................................................
      if(IPROC.NE.0) return   !  IPROC > 0,  TopRex process
* standard PYTHIA process is chosen, no TopRex call !
*..........................................................
         MSEL = 0           
         MSUB(16)=1         ! ffbar --> gW+ or gW-
*
      RETURN 
      END 
\end{verbatim} 
\noindent This example routine is used for further fast simulation
of CMS detector response (by call of {\tt CMSJET} routine) and for
writing of NTUPLE (by call of {\tt TX\_NT} routine).
\begin{verbatim} 
      SUBROUTINE USRPRO(MODE, NV)

      IMPLICIT NONE
      INTEGER MODE, NV         ! internal variables 
*...............................................................
      if(MODE.EQ.-1) THEN
*  initialisation
         call CMSJET(MODE)     ! reading and initialisation of CMSJET
      elseif(MODE.EQ.0) THEN   ! running   
         call CMSJET(MODE)     ! running of CMSJET
*
      elseif(MODE.EQ.1) then   ! closing files, etc     
        call CMSJET(MODE)    
      ENDIF
      END
\end{verbatim} 
\normalsize

\newpage
%%%%%%%%%%%%%%%%%%%%%%%%%%%%%%%%%%%%%%%%%%%%%%%%%%%%%%%%%%%%%%%%%

\section*{\bf Appendix~3.  }
\label{app3}

\vspace*{3mm}

\begin{center}
  {\bf Tools to read in parameters} 
\end{center}
The input data file {\tt run.dat} has a simple format.
The data card will be ignored, if it is: \\
$\bullet$ a totally blank card \\
$\bullet$ a card with the first non-blank '*' character (a comment card) \\
$\bullet$ a card without '=' character 

The first item in the card with data is a key of a variable.
The '=' character  should be explicitly given behind this key.
Then one or several numerical values have to be given.
The rest of the card is considered as comment.
Two examples of such cards are given below.
\begin{verbatim}
*   comment
 iproc  = 21        ! no. of TopRex process (21 = gg -> tt~ production)
 filout = myout.dat ! name of the output data file 
*    New Physics scale (1 TeV)    
 Rpar(60) = 1000.0 ! in GeV 
\end{verbatim}

In this example the items '{\tt iproc  =}', '{\tt filout =}' and 
'{\tt Rpar(60) =}' 
are the keys, while '21', 'myout.dat' and '1000.0' are the corresponding 
values to
be read. Four different subroutines are provided to read these data: 
\begin{verbatim}
   call TXRCHR(fkey, 'filout',     fname,     IER) ! character value
   call TXRINT(fkey, 'Iproc =',    iproc,     IER) ! integer value 
   call TXRFL8(fkey, 'rpar(60) =', Rpar(60),  IER) ! double precision value
   call TXRFL4(fkey, 'name =',     realvalue, IER) ! single precision value
\end{verbatim}
where the second parameter is a key-string, the third 
parameter is the value of the key variable,
and IER is an error flag ({\tt IER = 1} means that the subroutine is not able 
to find appropriate data in the file {\tt 'run.dat'}). 

The first parameter, {\tt fkey}, is the key for the file, from which
the parameters are to be read (the {\tt 'run.dat'} file in this example):
\begin{verbatim}
      character *6 fkey   ! key for file with input data 
      data fkey/'rundat'/
\end{verbatim}

The user can also read several numerical values from one card. As an example,
we present the card with the key of '{\tt H+- boson}', with two numerical
values, '220.' and '25.' being the mass of the $H^{\pm}$-boson and the 
$\tan\beta$ parameter. 
\begin{verbatim}
* mass and tan(beta) parameters for H+- boson process 
 H+- boson = 220.   25.   
\end{verbatim}

To read this card a character string (here {\tt STROUT}) should
 be described and the {\tt TXRSTR} subroutine is used to read these two 
values:
\begin{verbatim}
      CHARACTER *80 STROUT ! string to be used for data input
      ........
      call TXRSTR('h+- boson =', STROUT, IER) 
      read(STROUT,*) Rpar(51), Rpar(55)  ! mass and tan(beta)
      ........
\end{verbatim}

For string-key values in the {\tt run.dat} file and in
the parameters of the subroutines {\tt TXR...} lowercase 
and uppercase letters are allowed 
(the read in routines are not case sensitive).
Blank characters are allowed, too.

\newpage

%%%%%%%%%%%%%%%%%%%%%%%%%%%%%%%%%%%%%%%%%%%%%%%%%%%%%%%%%%%%%%%%%%%%%%%%%%%%%
\section*{\bf Appendix~4.}
\label{app4}
The parameters of the {\TTF /TXPAR/} common block are explained in the 
following. The default values of the parameters are
given in the parentheses (D = ...) 

\drawbox{COMMON/TXPAR/ Ipar(200), Rpar(200)}%
\begin{entry}

\itemc{Purpose:} to store information of several \TR global
 parameters.
\iteme{Ipar(1)    :} \TR's process number
\iteme{Ipar(2)    :} number of events to be generated 
\iteme{Ipar(4)    :} continuation flag 
\iteme{Ipar(5)    :} number of entries used for estimation of SigMax 
    (D = 200000)
\iteme{Ipar(6)    :} number of entries used for estimation of $|M|^2$
     (D = 100000)
\iteme{Ipar(7)   :} KF code of $Q$-quark for $W Q \bar Q$ process (D = 5,
                        $Q$=$b$-quark)
\iteme{Ipar(8)   :} quark masses, (D=1 : RPP values), 2 : \PYT values  
\iteme{Rpar(1)    :} \begin{math}\sqrt{\textstyle s}=E_{cm}\end{math} in GeV.
\iteme{Rpar(2)    :} $s = E_{cm}^2$ in GeV${}^2$.
\iteme{Rpar(3)    :} evolution scale for PDF evaluation 
\iteme{Rpar(4)    :} evolution scale for evaluation of $\alpha_s$  
\iteme{Rpar(10)   :} $\alpha_{QED} = (1/128)$ electromagnetic coupling 
\iteme{Rpar(11)   :} $e = \sqrt{4 \pi \alpha_{QED}}$ electric charge
\iteme{Rpar(12)   :} Fermi constant $G_F$
\iteme{Rpar(13:17):} $\sin \vartheta_W$,  $\cos \vartheta_W$, 
   $\sin^2 \vartheta_W$, $\cos \vartheta_W$, $\sin 2\vartheta_W$
\iteme{Rpar(18:19):} $g = e / \sin \vartheta_W$, $g_z = e /\sin 2\vartheta_W$
\iteme{Rpar(20)   :}  0.38939     transformation 1/GeV$^2$ to mb 
\iteme{Rpar(21:26):} $\pi$, $2 \pi$, $(2\pi)^3$, $(2\pi)^4$, $(2\pi)^6$
\iteme{Rpar(30)   :} SM top quark decay width ($t \to bW$)
\iteme{Rpar(31)   :} total top quark decay width (including non-SM 
                          interactions)
\iteme{Rpar(32)   :} partial top quark decay width for $t \to b H^{\pm}$
\iteme{Rpar(41:44):} $W$-boson parameters, $M_W$, $M^2_W$, $\Gamma(W)$,
                      $M_W \cdot \Gamma(W)$ 
\iteme{Rpar(45:48):} $Z$-boson parameters, $M_Z$, $M^2_Z$, $\Gamma(Z)$,
                      $M_Z \cdot \Gamma(Z)$ 
\iteme{Rpar(51:56):} $H^{\pm}$ boson parameters, $M_H$, $M^2_H$, 
     $\Gamma(H)$, $M \Gamma$, $\tan \beta$, $\cot \beta$
\iteme{Rpar(60)   : } (D=1000.d0) New Physics scale, $\Lambda = 1$~TeV
\iteme{Rpar(61:76):} anomalous FCNC top quark couplings 
({\tt FCNC\_INI} routine) 
%\pagebreak
\iteme{Rpar(81)   :} $g_{VV}$, relative coupling of Higgs boson to $W$ and $Z$
 vector bosons 
\iteme{\phantom{Rpar(81)   :}}  (processes Ipar(1) = 50, 51, 52 only)
\iteme{Rpar(82)   :} $a$, scalar Yukawa coupling of Higgs boson to top quarks
\iteme{\phantom{Rpar(82)   :}}  (processes Ipar(1) = 50, 51, 52 only)
\iteme{Rpar(83)   :} $\tilde{a}$, pseudoscalar Yukawa coupling of Higgs boson 
to top quarks 
\iteme{\phantom{Rpar(83)  :}}  (processes Ipar(1) = 50, 51, 52 only) 
\iteme{Rpar(101) :} $\sqrt{\hat s_{min}}$ for hard process 
                               (D = 20~GeV), identical to {\tt CKIN(1)} 
\iteme{Rpar(102) :} $\sqrt{\hat s_{max}}$ for hard process 
                              (D = $\sqrt{s}$), identical to {\tt CKIN(2)} 
\iteme{Rpar(103) :} minimal value for $\hat{p}_{\perp}$ 
                              (D = 5 GeV), identical to {\tt CKIN(3)}
\iteme{Rpar(104) :} minimal value for $\hat{p}_{\perp}$ 
                              (D = $\sqrt{s}/2$), identical to {\tt CKIN(4)}
\end{entry}

\newpage

%%%%%%%%%%%%%%%%%%%%%%%%%%%%%%%%%%%%%%%%%%%%%%%%%%%%%%%%%%%%%%%%%%%%%%%%%%%%%%%

\end{document}